\documentclass[prl,twocolumn,showpacs,superscriptaddress,floatfix]{revtex4}

\usepackage{amsmath}
\usepackage{graphicx}
\usepackage[dvips]{color}

\mathchardef\ogon="012C%
\newcommand{\as}{a\kern-0.22em\lower.40ex\hbox{$_{\ogon}$}}

\begin{document}
  \title{Spontaneous dissociation of long-range Fesh\-bach molecules}
  \author{Thorsten K\"{o}hler}
  \affiliation{Clarendon Laboratory, Department of Physics, 
    University of Oxford, Parks Road, Oxford, OX1 3PU, United Kingdom}
  \author{Eite Tiesinga}
  \author{Paul S.\ Julienne}
  \affiliation{Atomic Physics Division, National Institute of Standards and 
    Technology, 100 Bureau Drive Stop 8423, Gaithersburg, Maryland 20899-8423}
  
\begin{abstract}
  We study the spontaneous dissociation of diatomic molecules produced in 
  cold atomic gases via magnetically tunable Fesh\-bach resonances. 
  We provide a universal formula for the lifetime of these 
  molecules that relates their decay to the scattering length and the loss 
  rate constant for inelastic spin relaxation. Our universal 
  treatment as well as our exact coupled channels calculations for $^{85}$Rb 
  dimers predict a suppression of the decay over several orders of magnitude 
  when the scattering length is increased. Our predictions are in good 
  agreement with recent measurements of the lifetime of $^{85}$Rb$_2$.
\end{abstract}
\date{\today}
\maketitle

%\section{Introduction}
The discovery of Fesh\-bach resonances in cold gases provides the unique 
opportunity to widely tune the inter-atomic interactions via magnetic fields 
near a singularity of the scattering length. This singularity is due to the 
near degeneracy of the very small collision energy of the cold atoms with
the binding energy of an extremely loose diatomic molecular state for magnetic 
fields at which the scattering length is positive. A number of recent 
experiments, as summarised in the accompanying experimental paper of Thompson 
{\em et al.} \cite{Thompson04}, have taken advantage of this virtual energy 
match to populate this molecular state. 
The interacting pairs of atoms in the first studies of these cold Fesh\-bach 
molecules \cite{Donley02,Regal03} were subject to deeply inelastic spin 
relaxation collisions involving the deexcitation of at least one of 
the colliding atoms. Spin relaxation can affect both free \cite{Roberts00} 
and bound atom pairs. 

In this letter we predict the spontaneous dissociation 
of Fesh\-bach molecules due to spin relaxation. We provide a universal 
treatment of the molecular decay by which we
relate the molecular lifetime $\tau$ to the spatial extent of the bound state 
wave function and the loss rate constant for inelastic spin relaxation 
collisions. Both the size of the molecule and the loss rate constant exhibit
a pronounced resonance enhancement at the singularity of the scattering length.
General considerations on the nature of this enhancement allow us to predict 
the functional form of the magnetic field dependence of $\tau$. We show that 
the near resonant molecular lifetime increases with increasing scattering 
lengths and can therefore be varied over several orders of magnitude in related
experiments. We demonstrate the predictive power of our universal treatment in 
a comparison with exact coupled channels calculations of the lifetime of the 
$^{85}$Rb dimers produced in the experiments of 
Refs.~\cite{Donley02,Claussen03,Thompson04}. The exact calculations 
show that $\tau$ varies in the remarkable range between only 100 $\mu$s and 
roughly 30 ms in the experimental range of magnetic field strengths. Our 
studies indicate that it is the large spatial extent of the near resonant 
bound state wave function that can provide the crucial stability of the 
Fesh\-bach molecules with respect to their spontaneous decay. 
%We compare our 
%predictions with recent measurements of the lifetime of $^{85}$Rb dimers by 
%Thompson {\em et al.} \cite{Thompson04}. This comparison provides clear 
%evidence for the spontaneous dissociation of $^{85}$Rb$_2$ Fesh\-bach 
%molecules.

%\section{Analytic estimate of the molecular lifetime}
In the following, we shall consider pairs of atoms in a cold gas that 
interact via $s$ waves.
We denote the binary scattering channel of a pair of asymptotically 
free atoms as the entrance channel and choose the zero of energy at its 
dissociation threshold. This threshold is determined by the Zeeman 
energy of the separated atoms. 
%In the following we shall derive a general 
%expression for the molecular lifetime and relate it to the rate constant for 
%atom loss due to spin relaxation in an ultra-cold gas above the condensation 
%temperature. 
%We shall consider the region of magnetic field strengths on the 
%side of the zero energy resonance, where the scattering length is positive, 
The Hamiltonian $H$ of the relative motion can then
be divided into the contribution $H_\mathrm{cl}$ of the closed channels with 
dissociation thresholds at or above zero energy (including the entrance 
channel), the contribution $H_\mathrm{op}$ of the open channels with negative 
threshold energies, as well as the weak coupling $H_\mathrm{int}$ between the 
closed and open channels:
\begin{align}
  H=
  \left(
  \begin{array}{cc}
    H_\mathrm{op} & H_\mathrm{int}\\
    H_\mathrm{int}^\dagger & H_\mathrm{cl}
  \end{array}
  \right).
  \label{H2B}
\end{align}
 
We shall consider magnetic field strengths on the side of positive scattering 
lengths of a zero energy resonance (i.e.~a singularity of the scattering 
length). The large positive values of the scattering length imply the 
existence of a weakly bound (metastable) molecular state 
$\phi_{-1}^\mathrm{cl}$, which corresponds to the highest excited vibrational 
($v=-1$)
bound state of $H_\mathrm{cl}$. The wave function $\phi_{-1}^\mathrm{cl}$ thus 
fulfils the stationary Schr\"odinger equation 
$H_\mathrm{cl}\phi_{-1}^\mathrm{cl}=E_{-1}^\mathrm{cl}\phi_{-1}^\mathrm{cl}$, 
where $E_{-1}^\mathrm{cl}$ is the negative binding energy. 
%The probability for the decay of a Fesh\-bach molecule 
%at time $t$ is then given by: 
%\begin{align}
%  p_\mathrm{decay}(t-t_i)=1-
%  \left|
%  \langle\phi_{-1}^\mathrm{cl}|e^{-iH(t-t_i)/\hbar}
%  |\phi_{-1}^\mathrm{cl}\rangle
%  \right|^2.
%  \label{pdecay}
%\end{align}
%Here $t_i$ is the time at which the molecule was produced. As the transition 
%energy in a spin relaxation event stems from the atomic hyperfine states, it
%is much larger than $\left|E_{-1}^\mathrm{cl}\right|$ 
%(cf. Refs.~\cite{Donley02,Claussen03}) and the inter-channel coupling. The 
%probability $p_\mathrm{decay}(t-t_i)$ is then determined by Fermi's Golden 
%Rule in terms of an exponential decay:
%\begin{align}
%  \left|
%  \langle\phi_{-1}^\mathrm{cl}|e^{-iH(t-t_i)/\hbar}
%  |\phi_{-1}^\mathrm{cl}\rangle
%  \right|^2=e^{-\gamma(t-t_i)}.
%\end{align}
Fermi's golden rule determines the exponential decay of the Fesh\-bach 
molecule by a rate
\begin{align} 
  \gamma=-\frac{2}{\hbar} \ \mathrm{Im}
  \left[
  \langle\phi_{-1}^\mathrm{cl}|
  H_\mathrm{int}^\dagger G_\mathrm{op}(E_{-1}^\mathrm{cl}+i0)H_\mathrm{int}
  |\phi_{-1}^\mathrm{cl}\rangle
  \right].
  \label{gamma}
\end{align}
Here $G_\mathrm{op}(E_{-1}^\mathrm{cl}+i0)=
\left(E_{-1}^\mathrm{cl}+i0-H_\mathrm{op} \right)^{-1}$
is the Green's function of the open binary scattering channels,
whose energy argument $E_{-1}^\mathrm{cl}+i0$ indicates that 
$E_{-1}^\mathrm{cl}$ is approached from the upper half of the complex plane. 
As $\left|E_{-1}^\mathrm{cl}\right|$ is much smaller than typical transition 
energies, it can be neglected in the argument of the Green's function. 
%The lifetime of the Fesh\-bach molecules is given by $\tau=1/\gamma$.  

Fermi's golden rule also determines the loss rate constant for 
inelastic spin relaxation collisions to be
\begin{align} 
  K_2=-\frac{4}{\hbar}(2\pi\hbar)^3 \ \mathrm{Im}
  \left[
    \langle\phi_{0,\mathrm{cl}}^{(+)}|
    H_\mathrm{int}^\dagger G_\mathrm{op}(i0)H_\mathrm{int}
    |\phi_{0,\mathrm{cl}}^{(+)}\rangle
    \right].
  \label{K2}
\end{align}
Here $\phi_{0,\mathrm{cl}}^{(+)}$ is the zero energy scattering state 
associated with the Hamiltonian $H_\mathrm{cl}$, which fulfils the stationary 
Schr\"odinger equation $H_\mathrm{cl}\phi_{0,\mathrm{cl}}^{(+)}=0$. The 
prefactors in Eq.~(\ref{K2}) correspond to a cold thermal gas of identical 
Bosons. The associated rate equation reads \cite{Roberts00} 
$\dot{N}(t)=-K_2\langle n(t)\rangle N(t)$, where $N(t)$ is 
the number of atoms and $\langle n(t)\rangle$ is their average density.

%The right hand sides of Eqs.~(\ref{gamma}) and (\ref{K2}) involve expectation
%values of the same matrix in different energy states of $H_\mathrm{cl}$. 
%We shall show 
%in the following that the zero energy scattering state 
%$\phi_{0,\mathrm{cl}}^{(+)}$ 
%and the vibrational bound state $\phi_{-1}^\mathrm{cl}$ have essentially the 
%same functional form on the length scales set by the range of the interchannel
%coupling potential $H_\mathrm{int}$. The proportionality factor between 
%$\phi_{0,\mathrm{cl}}^{(+)}$ and $\phi_{-1}^\mathrm{cl}$ at short inter-atomic
%distances relates $K_2$ to $\tau$. 
%The derivation of this proportionality factor at near resonant
%magnetic field strengths will be the objective of the following 
%considerations.
 
At magnetic field strengths in the vicinity of a zero energy resonance, 
$\phi_{0,\mathrm{cl}}^{(+)}$ and $\phi_{-1}^\mathrm{cl}$ are determined by 
their wave functions in 
the entrance channel and in a closed channel strongly coupled to it
\cite{Mies00,Goral03}. This strong inter-channel coupling is due to the near 
degeneracy of the energy $E_\mathrm{res}(B)$ of a closed channel vibrational 
state (the bare Fesh\-bach resonance level) $\phi_\mathrm{res}$ with the 
dissociation threshold of the entrance channel. The restricted 
two-body Hamiltonian $H_\mathrm{cl}$ of the entrance channel and the closed 
channel can be effectively described by:  
\begin{align}
  H_\mathrm{cl}=
  \left(
  \begin{array}{cc}
    -\frac{\hbar^2}{m}\nabla^2+V_\mathrm{bg}(r) & W(r)\\
    W(r) & -\frac{\hbar^2}{m}\nabla^2+V_\mathrm{cl}(B,r)
  \end{array}
  \right).
  \label{H>}
\end{align}
Here $r$ denotes the relative distance between the atoms and $m$ is twice their
reduced mass, $B$ is the magnetic field strength, $V_\mathrm{bg}(r)$ is the 
background scattering potential of the entrance channel, and $W(r)$ provides 
the inter-channel coupling. The closed channel potential $V_\mathrm{cl}(B,r)$ 
supports the resonance state, 
i.e.~$[-\hbar^2\nabla^2/m+V_\mathrm{cl}(B,r)]\phi_\mathrm{res}(r)=
E_\mathrm{res}(B)\phi_\mathrm{res}(r)$. 
The dissociation threshold of $V_\mathrm{cl}(B,r)$ is determined by the energy 
of a pair of non-interacting atoms in the closed channel. The relative Zeeman 
energy shift between the two channels as well as the bare energy 
$E_\mathrm{res}(B)$ 
can be tuned by varying the magnetic field strength $B$. We denote by 
$B_\mathrm{res}$ the magnetic field strength, at which $E_\mathrm{res}(B)$ 
crosses the dissociation threshold energy of the entrance channel, 
i.e.~$E_\mathrm{res}(B_\mathrm{res})=0$. 
%In the restricted range of 
%magnetic field strengths in the vicinity of the zero energy resonance 
The bare energy $E_\mathrm{res}(B)$ varies virtually linearly in $B$, and an 
expansion about $B_\mathrm{res}$ yields
%\begin{equation}
$E_\mathrm{res}(B)=\mu_\mathrm{res}(B-B_\mathrm{res})$.
%  \label{slope}
%\end{equation}
Here $\mu_\mathrm{res}$ is the difference in magnetic moment between the 
Fesh\-bach resonance state and a pair of atoms in the entrance channel.

%Reference \cite{Goral03} provides a detailed description of all two component
%energy states of the Hamiltonian $H_\mathrm{cl}$ in Eq.~(\ref{H\mathrm{cl}}) 
%and relates the 
%potentials to measurable parameters of a Fesh\-bach resonance. We shall 
%therefore just quote those results that are needed for the present 
%considerations. 

Under the assumption that the spatial configuration of a pair of atoms in the 
closed channel is restricted to the resonance state 
$\phi_\mathrm{res}(r)$, the dressed highest excited vibrational bound state 
$\phi_{-1}^\mathrm{cl}$ of $H_\mathrm{cl}$ is given by 
\cite{Goral03}:
\begin{align}
  \phi_{-1}^\mathrm{cl}
  =\frac{1}{\mathcal{N}}
  \left(
  \begin{array}{c}
    G_\mathrm{bg}(E_{-1}^\mathrm{cl})W\phi_\mathrm{res}\\
    \phi_\mathrm{res}
  \end{array}
  \right).
  \label{phiminus1}
\end{align}
Here $G_\mathrm{bg}(E_{-1}^\mathrm{cl})=
(E_{-1}^\mathrm{cl}+\hbar^2\nabla^2/m-V_\mathrm{bg})^{-1}$
is the Green's function associated with the entrance channel, and the factor 
$1/\mathcal{N}$ assures that $\phi_{-1}^\mathrm{cl}$ is unit normalised. The 
associated binding energy $E_{-1}^\mathrm{cl}$ is determined by \cite{Goral03}
$E_{-1}^\mathrm{cl}=E_\mathrm{res}(B)+\langle\phi_\mathrm{res}|
WG_\mathrm{bg}(E_{-1}^\mathrm{cl})W|\phi_\mathrm{res}\rangle$.
The strong coupling between the entrance channel and the bare Fesh\-bach 
resonance state $\phi_\mathrm{res}(r)$ also determines the zero energy 
scattering state $\phi_{0,\mathrm{cl}}^{(+)}$ of $H_\mathrm{cl}$ by the 
general formula \cite{Goral03}:
\begin{align}
  \phi_{0,\mathrm{cl}}^{(+)}
  =
  \left(
  \begin{array}{c}
    \phi_{0,\mathrm{bg}}^{(+)}+G_\mathrm{bg}(0)W\phi_\mathrm{res} \ A(B)\\
    \phi_\mathrm{res} \ A(B)
  \end{array}
  \right).
  \label{phi0>}
\end{align}
Here $\phi_{0,\mathrm{bg}}^{(+)}$ denotes the bare zero energy state 
associated with the background scattering, 
i.e.~$[-\hbar^2\nabla^2/m+V_\mathrm{bg}(r)]\phi_{0,\mathrm{bg}}^{(+)}(r)=0$, 
and the amplitude $A(B)$ in Eq.~(\ref{phi0>}) is given by \cite{Goral03}: 
\begin{align}
  A(B)=-\frac{\langle\phi_\mathrm{res}|W|\phi_{0,\mathrm{bg}}^{(+)}\rangle}
  {E_\mathrm{res}(B)+
    \langle\phi_\mathrm{res}|WG_\mathrm{bg}(0)W|\phi_\mathrm{res}\rangle}.
  \label{amplitude}
\end{align}
The scattering length $a(B)$ can then be obtained from the 
asymptotic behaviour  
$(2\pi\hbar)^{-3/2}[1-a(B)/r]$ of the entrance channel component  
of Eq.~(\ref{phi0>}) at large inter-atomic distances $r$. This yields
\begin{align}
  a(B)=a_\mathrm{bg}\left(1-\frac{\Delta B}{B-B_0}\right),
  \label{aofB}
\end{align}
where 
$\Delta B=(2\pi\hbar)^3|\langle\phi_\mathrm{res}|W|\phi_{0,\mathrm{bg}}^{(+)}
\rangle|^2 m/(4\pi\hbar^2 a_\mathrm{bg}\mu_\mathrm{res})$
is the resonance width and $a_\mathrm{bg}$ is the background scattering 
length. The singularity of $A(B)$ thus determines the 
measurable position $B_0$ of the zero energy resonance, which is distinct from 
the zero $B_\mathrm{res}$ of the bare energy $E_\mathrm{res}(B)$ by the
shift $B_0-B_\mathrm{res}=-\langle\phi_\mathrm{res}|WG_\mathrm{bg}(0)W
|\phi_\mathrm{res}\rangle/\mu_\mathrm{res}$ of the denominator on the right 
hand side of Eq.~(\ref{amplitude}).

%The asymptotic form of the component of $\phi_{0,\mathrm{bg}}^{(+)}(r)$ in 
%the entrance channel at large interatomic separations determines the well 
%known formula 
%\begin{align}
%a(B)=a_\mathrm{bg}\left(1-\frac{\Delta B}{B-B_0}\right)
%\label{aofB}
%\end{align}
%for the magnetic field dependence of the scattering length \cite{Goral03}. 
%Here $B_0$ is the measurable resonance position, i.e.~the magnetic field 
%strength at which the scattering length has a singularity, and $\Delta B$ 
%is usually referred to as the resonance width. The resonance position is 
%identical to the zero of the resonance denominator of the amplitude $A(B)$ 
%in Eq.~(\ref{amplitude}). We note that $B_0$ is shifted with respect to the 
%magnetic field strength $B_\mathrm{res}$ at which the resonance state energy 
%$E_\mathrm{res}(B)$ crosses the dissociation threshold of the entrance 
%channel. In the case of the 155 Gauss Fesh\-bach resonance of $^{85}$Rb this 
%shift is as large as $B_0-B_\mathrm{res}=-9$ Gauss.

At near resonant magnetic field strengths the binding energy is determined by 
$E_{-1}^\mathrm{cl}=-\hbar^2/(ma^2)$ 
(cf., e.g., Refs.~\cite{Donley02,Claussen03,Goral03}). The 
energy argument of the Green's function in Eq.~(\ref{phiminus1}), therefore, 
vanishes at large positive scattering lengths. In this limit the background 
scattering contribution $\phi_{0,\mathrm{bg}}^{(+)}$ to the right hand side of 
Eq.~(\ref{phi0>}) also becomes negligible for inter-atomic distances on the 
length scale set by the range of the inter-channel coupling $H_\mathrm{int}$ 
in Eqs.~(\ref{gamma}) and (\ref{K2}). Equations (\ref{phiminus1}) and 
(\ref{phi0>}) thus coincide up to a factor $\mathcal{N}A(B)$, and
Eqs.~(\ref{gamma}) and (\ref{K2}) yield  
$K_2(B)/\gamma(B)=2(2\pi\hbar)^3|A(B)|^2 \mathcal{N}^2(B)$. The near resonant 
amplitude $A(B)$ is readily obtained from the resonance width and shift 
of Eq.~(\ref{aofB}), while the normalisation constant 
$\mathcal{N}^2(B)=(\mu_\mathrm{res}\Delta B) m a_\mathrm{bg} 
a(B)/(2\hbar^2)$ can be determined from  
$\langle\phi_{-1}^\mathrm{cl}|\phi_{-1}^\mathrm{cl}\rangle=1$ \cite{Goral03}. 
The molecular lifetime $\tau(B)=1/\gamma(B)$ is then given by the
universal formula 
\begin{align}
  \tau(B)=4\pi a^3(B)/K_2(B)=
  [4/K_2(B)]4\pi\langle r^3\rangle/3.
  \label{tauofBuniv}
\end{align}
Here $\langle r^3\rangle$ is the expectation value of $r^3$ for the
near resonant wave function 
$\phi_{-1}^\mathrm{cl}(r)=\exp(-r/a)/(r\sqrt{2\pi a})$ 
\cite{TKTGPSJKB03,Goral03}.
%In this formula $a^3(B)$ sets the scale of the 
%volume that confines the bound atom pair \cite{TKTGPSJKB03}, and $K_2(B)$
%provides the rate constant for the inelastic collision event. 
%Both $a(B)$ 
%and $K_2(B)$ are measurable quantities \cite{Roberts00}. 

%\section{Resonance interpretation}
%The determination of the magnetic field dependence of $K_2(B)$ in 
%Eq.~(\ref{tauofBuniv}) provides an experimental challenge \cite{Roberts00}. 
General considerations \cite{Bohn} show that in the presence of 
open decay channels the resonance enhanced zero energy binary elastic and 
inelastic collision cross sections can be described in terms of a complex 
scattering length $a(B)-ib(B)$, whose imaginary part is related to the 
loss rate constant by 
\begin{align}
  K_2(B)=16\pi\hbar b(B)/m.
  \label{K2ofB}
\end{align}
Its real part is well approximated by Eq.~(\ref{aofB}), except that $a^2(B)$ 
approaches a large but finite value at $B=B_0$, which is determined by the 
decay width. We shall presuppose in the following that only the bare resonance 
state $\phi_\mathrm{res}$ decays to a set of final channels with a total rate 
$\gamma_\mathrm{res}=1/\tau_\mathrm{res}$, where $\tau_\mathrm{res}$ is 
the associated lifetime. The imaginary part of the complex scattering length 
is then given by a Lorentzian function
\begin{align}
  %a=& a_{bg} - a_{bg} 
  %\Delta B \frac{B-B_0}{(B-B_0)^2+
  %  [\gamma_\mathrm{res}/(4\mu_\mathrm{res})]^2} \label{ares1}\\
  %&=& a_{bg} \left ( 1 - \frac{\Delta_n}{B-B_0} \right )  \,\,\, 
  %\mathrm{if} {\gamma'}_\mathrm{res}=0 \label{ares2}\\
  b(B)=a_{bg}\Delta B 
  \frac{\hbar\gamma_\mathrm{res}/(4\mu_\mathrm{res})}{(B-B_0)^2+ 
    [\hbar\gamma_\mathrm{res}/(4\mu_\mathrm{res})]^2}. 
  \label{bofB}
\end{align}
Inserting Eqs.~(\ref{aofB}), (\ref{K2ofB}) and (\ref{bofB}) into 
Eq.~(\ref{tauofBuniv}) then yields 
\begin{align}
  \tau(B)=\tau_\mathrm{res} 
  \frac{ma_\mathrm{bg}^2\mu_\mathrm{res}\Delta B}{4 \hbar^2} x^2 
  \left(\frac{1}{x}-1\right)^3=\tau_\mathrm{res} f(B),
  \label{tauB}
\end{align}
where we have introduced $x=(B-B_0)/\Delta B$.
This formula provides the general functional form of $\tau(B)$ as well as its
order of magnitude in terms of the constant factor $\tau_\mathrm{res}$,
which can be obtained either from a measurement of $K_2(B)$, at a single 
magnetic field strength, or from coupled channels calculations.

We shall demonstrate the predictive power of 
Eqs.~(\ref{tauofBuniv}) and (\ref{tauB}) for the example of the Fesh\-bach 
molecules associated recently from pairs of $^{85}$Rb atoms in  
a Bose-Einstein condensate \cite{Donley02,Claussen03,Thompson04} at magnetic 
field strengths of about 15.5 mT. The nuclear spin $I=5/2$ of $^{85}$Rb gives 
rise to two energetically relevant hyperfine levels with total spin quantum 
numbers $f=2$ and 3. A homogeneous magnetic field $B$ splits the Zeeman 
sublevels into a non-degenerate set of levels $(f,m_f)$, labelled by the 
projection quantum number $m_f$ along the axis of the magnetic field and the 
$f$ value with which it correlates adiabatically at zero field. In the 
experiments \cite{Donley02,Claussen03} the atoms were prepared in the 
$(f=2,m_f=-2)$ internal state. The binary asymptotic scattering channels are 
then characterised by pairs of internal quantum numbers $(f,m_f;f',m_f')$, as 
well as the orbital angular momentum quantum number $\ell$ associated with the 
relative motion of the atom pair and its projection quantum number $m_\ell$. 
The $s$ wave $(\ell=0,m_\ell=0)$ entrance channel is thus characterised just 
by the internal quantum numbers $(2,-2;2,-2)$. The cylindrical symmetry of the
setup implies that the projection quantum number $m_f+m_f'+m_\ell=-4$ of the 
total angular momentum is conserved in a binary collision.  

\begin{figure}[htb] 
  \includegraphics[width=\columnwidth,clip]{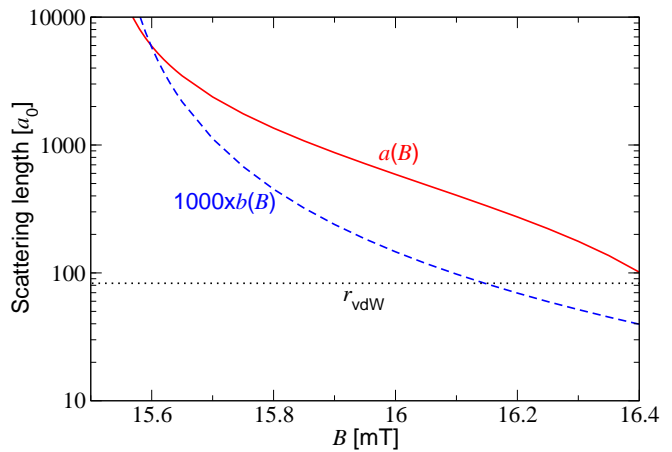}
  \caption{Magnetic field dependence of the real (solid curve) and imaginary 
    (dashed curve) parts of the 
    complex scattering length $a(B)-ib(B)$. The dotted line indicates the 
    length scale \cite{GribakinFlambaum93} 
    $r_\mathrm{vdW}=(m C_6/\hbar^2)^{1/4}/2$ set 
    by the long range asymptotic $-C_6/r^6$ van der Waals tail of the 
    background scattering potential. We note that $b(B)$ is multiplied by a 
    factor of 1000. \label{fig:a-ib}}
\end{figure}

We have applied the coupled channels approach of Ref.~\cite{Mies00} to 
determine the scattering length $a(B)$ (see Fig.~\ref{fig:a-ib}), including 
spin exchange interactions, which couple the entrance channel to the four 
closed $s$ wave scattering channels $(3,-3;2,-1)$, $(3,-2;2,-2)$, 
$(3,-1;3,-3)$ and $(3,-2;3,-2)$. A fit to Eq.~(\ref{aofB}) predicts the 
parameters $B_0=$ 15.520 mT, $\Delta B=$ 1.065 mT, and 
$a_\mathrm{bg}=$ -484.1 $a_0$ ($a_0$ = 0.0529177 nm) for the experimentally 
well characterised 15.5 mT zero energy resonance of $^{85}$Rb 
%\cite{Claussen03,Marte02}. 
\cite{Claussen03}.
The decay of the Fesh\-bach molecules is due to 
spin-dipole interactions, which couple $s$ waves to $d$ waves. Among the 23 
$d$ wave channels coupled to the entrance channel only three are open. These 
$d$ wave decay exit channels are characterised by the internal quantum numbers 
$(2,-2;2,-1)$, $(2,-2;2,0)$ and $(2,-1;2,-1)$ in order of increasing energy
release. The transition energies are all on the order of several mK
(in units of $k_\mathrm{B}=1.380 6505\times 10^{-23}$ J/K), which implies that 
the decay products are lost from an atom trap. Including spin-dipole 
interactions in the coupled channels calculations shifts the predicted 
position $B_0$ by less than 1 $\mu$T but it provides the small imaginary part 
$b(B)$ of the scattering length (see Fig.~\ref{fig:a-ib}), which determines 
the loss rate constant $K_2(B)$ by Eq.~(\ref{K2ofB}).

\begin{figure}[htb] 
  \includegraphics[width=\columnwidth,clip]{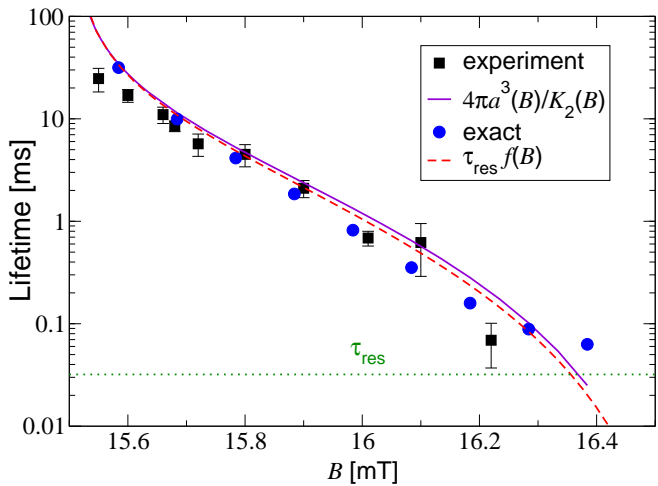}
  \caption{Two-body decay lifetime $\tau(B)$ of the $^{85}$Rb$_2$ 
    Fesh\-bach molecules \cite{Donley02,Claussen03} versus magnetic field 
    strength $B$ near 15.5 mT. The squares indicate the experimental data 
    of Thompson {\em et al.} \cite{Thompson04}. The circles indicate exact 
    predictions of coupled channels calculations, whereas the solid and dotted 
    curves show the results of the universal formulae (\ref{tauofBuniv}) and 
    (\ref{tauB}), respectively. The magnitude of the bare state lifetime 
    $\tau_\mathrm{res}$ is indicated by the dotted line. For the purpose of 
    comparison, we have slightly corrected the axis of the magnetic field 
    strength in such a way that the theoretical resonance position recovers 
    the measured value of $B_0=15.5041$ mT \cite{Claussen03}.
    \label{fig:lifetimes}}
\end{figure}

Given the magnetic moment difference $\mu_\mathrm{res}/h = 34.66$ MHz/mT, a fit
of Eq.~(\ref{bofB}) to the curve of $b(B)$ in Fig.~\ref{fig:a-ib} determines 
the lifetime of the bare resonance level to be 
$\tau_\mathrm{res}=1/\gamma_\mathrm{res}=32$ $\mu$s. The solid curve in 
Fig.~\ref{fig:lifetimes} shows the associated universal estimates of the 
lifetimes $\tau(B)$ of the dressed $^{85}$Rb$_2$ Fesh\-bach molecules 
\cite{Donley02,Claussen03} 
determined from Eq.~(\ref{tauofBuniv}), using the exact loss rate constant
$K_2(B)$ [cf.~Eq.~(\ref{K2ofB}) and Fig.~\ref{fig:a-ib}]. The 
dashed curve indicates the predictions of Eq.~(\ref{tauB}). 
We expect these estimates to be accurate between $B_0$ and about 16.0 mT, 
where the scattering length $a(B)$ by far exceeds the length scale 
$r_\mathrm{vdW}=84\, a_0$ set by the long range asymptotic behaviour of the 
van der Waals interaction (see Fig.~\ref{fig:a-ib}). This trend is confirmed 
by our direct calculations of the molecular lifetime shown in 
Fig.~\ref{fig:lifetimes}. The exact lifetime can be obtained from the energy 
dependence of the inelastic scattering cross section between any pair of open 
decay channels. We have chosen transitions between the open $d$ wave 
channels associated with the internal atomic quantum numbers $(2,-2;2,-1)$ and 
$(2,-2;2,0)$. Fitting the standard Breit-Wigner formula to the resonance peak 
at the binding energy $E_\mathrm{b}(B)$ \cite{Donley02,Claussen03} of the 
metastable Fesh\-bach molecules then determines the decay rate 
$\gamma(B)=1/\tau(B)$ through its width. 

The squares in Fig.~\ref{fig:lifetimes} indicate lifetime measurements by 
Thompson {\em et al.} \cite{Thompson04} in which the dimer molecules were 
produced in a dilute vapour of $^{85}$Rb atoms at a temperature of 30 nK using 
a Fesh\-bach resonance crossing technique 
(see, e.g., Refs.~\cite{Mies00,Goral03}). 
%These measurements were performed at low densities 
%($6.6\times 10^{11}\,$cm$^{-3}$) to largely rule out losses by inelastic 
%atom-molecule or molecule-molecule collisions.   
Our theoretical predictions agree with the experimental lifetimes in 
both their systematic trends and magnitudes. Figure \ref{fig:lifetimes} 
shows the large differences in the order of magnitude of $\tau(B)$ from 
roughly 30 ms to only 100 $\mu$s in the experimentally accessible 
\cite{Thompson04} range of magnetic field strengths from about 15.6 to 
16.2 mT. The lifetime correlates 
with the admixture of the bare resonance level to the dressed molecular state, 
which, in turn, is related to the spatial extent of the molecules 
\cite{TKTGPSJKB03}. Equation (\ref{tauofBuniv}) can indeed be interpreted in 
the intuitive way that $4\pi\langle r^3\rangle/3$ sets the scale of the volume 
confining the bound atom pair, while $K_2(B)/4$ is the event rate constant for 
molecular decay (which is the same as that for the inelastic collision of a 
pair of identical Bose condensed atoms \cite{Stoof89}). The large extent 
of the molecular wave function at near resonant magnetic field strengths thus 
protects the Fesh\-bach molecules against their spontaneous decay. 
We caution, however, that purely two-body theory may eventually become 
inadequate sufficiently close to resonance; for example, the mean distance 
between atoms at the experimental \cite{Thompson04} peak density of 
$6.6\times 10^{11}\,$cm$^{-3}$ is only $2\times 10^4$ $a_0$, which is 
comparable to the estimated bond length of the molecules \cite{TKTGPSJKB03} 
of about $3\times 10^3$ $a_0$ at 15.6 mT. 

Our general theory applies to a variety of 
%other experimentally relevant 
Fesh\-bach molecules with open decay exit channels \cite{Fermions}. Among 
these species are the $^{40}$K$_2$ \cite{Regal03} and $^6$Li$_2$ 
\cite{Bartenstein04} 
molecules produced in 
%two component 
Fermi gases 
%using a Fesh\-bach resonance crossing technique 
in the $s$ wave entrance channels 
associated with the pairs of atomic quantum numbers $(9/2,-9/2;9/2,-5/2)$ and 
$(1/2,1/2;3/2,-3/2)$, respectively. 
%For example, Ref.~\cite{Bartenstein04} 
%applied Eq.~(\ref{tauofBuniv}) to predict a purely two-body molecular 
%lifetime in the latter $^6$Li$_2$ case of greater than 10 s at 60.0 mT, 
%increasing to 1000 s at 68.5 mT very close to resonance at 69.04 mT.   
Measurements of two-body decay lifetimes 
could allow the characterisation of the spin-dipole interaction between various
atomic species with an unprecedented precision. Our universal estimate 
%provided in this letter 
determines the conditions for the stability of 
Fesh\-bach molecules, which is crucial to all future studies of cold molecular 
gases in the presence of open decay exit channels. 

We are grateful to Eleanor Hodby, Sarah Thompson and Carl Wieman
for providing their data prior to publication. This research has been 
supported by the Royal Society (T.K.) 
and the US Office of Naval Research (E.T. and P.S.J.).


\begin{thebibliography}{99}
\bibitem{Thompson04}
  S.T. Thompson, E. Hodby, and C.E. Wieman, cond-mat/0408144.
\bibitem{Donley02}
  E.A. Donley {\em et al.}, Nature (London)
  \textbf{417}, 529 (2002).
\bibitem{Regal03}
  C.A. Regal {\em et al.}, Nature (London)
  \textbf{424}, 47 (2003).
\bibitem{Roberts00}
  J.L. Roberts {\em et al.},
  Phys. Rev. Lett. \textbf{85}, 728 (2000).
\bibitem{Claussen03}
  N.R. Claussen {\em et al.}, Phys. Rev. A \textbf{67}, 060701(R) (2003).
\bibitem{Mies00}
  F.H. Mies, E. Tiesinga, and P.S. Julienne,
  Phys. Rev. A \textbf{61}, 022721 (2000).
\bibitem{Goral03}
  K. Goral {\em et al.},
  %cond-mat/0312178.
  J. Phys. B \textbf{37}, 3457 (2004).
\bibitem{TKTGPSJKB03}
  T. K\"ohler {\em et al.},
  Phys. Rev. Lett. \textbf{91}, 230401 (2003).
\bibitem{Bohn}
  J.L. Bohn and P.S. Julienne, Phys. Rev. A {\bf 56}, 1486 (1997);
  {\em ibid}.~{\bf 60}, 414 (1999).
%\bibitem{Julienne02}
% P.S.~Julienne, {\em Ultra cold collisions of atoms and
%   molecules}, Chapter 2.6.3 in {\em Scattering: Scattering and Inverse
%   Scattering in Pure and Applied Science}, edited by R. Pike and P. Sabatier
% (Academic Press, 2002), pp. 1043-1067.
\bibitem{GribakinFlambaum93}
  G.F. Gribakin and V.V. Flambaum, Phys. Rev. A \textbf{48}, 546 (1993).
%\bibitem{Marte02}
%  While our potentials have not been optimised based on the Fesh\-bach 
%  resonance data from A. Marte {\em et al.} 
%  [Phys. Rev. Lett.~\textbf{89}, 283202 (2002)],
%  we predict their positions $B_0$ to an accuracy of about 1\%.
\bibitem{Stoof89}
  H.T.C. Stoof {\em et al.}, Phys. Rev. A \textbf{39}, 3157 (1989).
\bibitem{Fermions}
  In the case of molecules composed of Fermions in different spin states the 
  event rate constant
  $K_2/4$ in Eq.~(\ref{tauofBuniv}) needs to be replaced by $K_2/2$.
\bibitem{Bartenstein04}
  Based on the results of this letter, the lifetimes of $^6$Li$_2$ Fesh\-bach 
  molecules were calculated by M. Bartenstein {\em et al.}, cond-mat/0408673.
  %A. Altmeyer, R. Geursen, S. Jochim, C. Chin, J. Hecker-Denschlag, 
  %R. Grimm, A. Simoni, E. Tiesinga, C.J. Williams, and 
  %P.S. Julienne 
%\bibitem{Thompson04}
%  S.T. Thompson, E. Hodby, and C.E. Wieman, cond-mat/0408144.
%\bibitem{Donley02}
%  E.A. Donley, N.R. Claussen, S.T. Thompson, and C.E. Wieman, Nature (London) 
% \textbf{417}, 529 (2002).
%\bibitem{Regal03}
%  C.A.~Regal, C.~Ticknor, J.L.~Bohn, and D.S.~Jin, Nature (London)
%  \textbf{424}, 47 (2003).
%\bibitem{Roberts00}  
%  J.L. Roberts, N.R. Claussen, S.L. Cornish, and C.E. Wieman,
% Phys. Rev. Lett. \textbf{85}, 728 (2000).
%\bibitem{Claussen03}
% N.R. Claussen, S.J.J.M.F. Kokkelmans, S.T. Thompson, E.A. Donley,
%  E. Hodby, and C.E. Wieman, Phys. Rev. A \textbf{67}, 060701 (2003).
%\bibitem{Mies00} 
% F.H. Mies, E. Tiesinga, and P.S. Julienne, 
% Phys. Rev. A \textbf{61}, 022721 (2000).
%\bibitem{Goral03}
%   K. Goral, T. K\"ohler, S.A. Gardiner, E. Tiesinga, and P.S. Julienne, 
%   cond-mat/0312178.
%\bibitem{TKTGPSJKB03}
% T.~K\"ohler, T.~Gasenzer, P.S.~Julienne, and K.~Burnett, 
%  Phys. Rev.~Lett.~\textbf{91}, 230401 (2003).
%\bibitem{Bohn}
%  J. Bohn and P.S. Julienne, Phys. Rev. A {\bf 56}, 1486 (1997);
% {\em ibid}.~{\bf 60}, 414 (1999).
%\bibitem{Julienne02} 
%  P.S.~Julienne, {\em Ultra cold collisions of atoms and 
%    molecules}, Chapter 2.6.3 in {\em Scattering: Scattering and Inverse 
%    Scattering in Pure and Applied Science}, edited by R. Pike and P. Sabatier
%  (Academic Press, 2002), pp. 1043-1067.
%\bibitem{GribakinFlambaum93}
%  G.F.~Gribakin and V.V.~Flambaum, Phys.~Rev.~A \textbf{48}, 546 (1993).
%\bibitem{Marte02}
%  While our potentials have not been optimised based on the Fesh\-bach 
%  resonance
%  data from A. Marte, T. Volz, J. Schuster, S. D\"urr, G. Rempe, E.G.M. van 
%  Kempen, and B.J. Verhaar [Phys.~Rev.~Lett.~\textbf{89}, 283202 (2002)],
%  we predict their positions $B_0$ to an accuracy of about 1\%.
%\bibitem{Stoof89}
%H.T.C. Stoof {\em et al.}, Phys.~Rev.~A \textbf{39}, 3157 (1989).
%\bibitem{Fermions}
%In the case of molecules composed of Fermions in different spin states the 
%event rate constant 
%$K_2/4$ in Eq.~(\ref{tauofBuniv}) needs to be replaced by $K_2/2$.  
%\bibitem{Bartenstein04}
%  M. Bartenstein, A. Altmeyer, R. Geursen, S. Jochim, C.~Chin, 
%  J. Hecker-Denschlag, R. Grimm, A. %Simoni, E. Tiesinga, C.J. Williams, 
%  and P.S. Julienne (unpublished).
\end{thebibliography}
\end{document}